\documentclass[tighten,twocolumn, twocolappendix]{aastex631}
\shorttitle{EUV LATE PHASE IN A FLARE}
\shortauthors{Li et al.}
\usepackage{amsmath}
\usepackage{apjfonts}

\begin{document}

\title{Enhanced Peak and Extended Cooling of the Extreme-ultraviolet Late Phase in a Confined Solar Flare}

\author{Shihan~Li}
\affiliation{School of Astronomy and Space Science, Nanjing University, Nanjing 210023, People's Republic of China}

\author{Yu~Dai}
\affiliation{School of Astronomy and Space Science, Nanjing University, Nanjing 210023, People's Republic of China}
\affiliation{Key Laboratory of Modern Astronomy and Astrophysics (Nanjing University), Ministry of Education, Nanjing 210023, People's Republic of China}
		
\author{Mingde~Ding}
\affiliation{School of Astronomy and Space Science, Nanjing University, Nanjing 210023, People's Republic of China}
\affiliation{Key Laboratory of Modern Astronomy and Astrophysics (Nanjing University), Ministry of Education, Nanjing 210023, People's Republic of China}

\author{Jinhan~Guo}
\affiliation{School of Astronomy and Space Science, Nanjing University, Nanjing 210023, People's Republic of China}
\affiliation{Centre for Mathematical Plasma Astrophysics, Department of Mathematics, KU Leuven, Celestijnenlaan 200B, B-3001 Leuven, Belgium}

\author{Hao~Wu}
\affiliation{School of Astronomy and Space Science, Nanjing University, Nanjing 210023, People's Republic of China}

\correspondingauthor{Yu~Dai}
\email{ydai@nju.edu.cn}

\begin{abstract}
We present observations and analysis of an X1.8 non-eruptive solar flare on 2012 October 23, which is characterized by an extremely large late-phase peak seen in the warm coronal extreme-ultraviolet (EUV) emissions ($\sim$ 3 MK), with the peak intensity over 1.4 times that of main flare peak. The flare is driven by a failed eruption of a magnetic flux rope (MFR), whose strong squeeze force acting on the overlying magnetic structures gives rise to an intense early heating of the late-phase loops. Based on differential emission measure (DEM) analysis, it is found that the late-phase loops experience a ``longer-than-expected" cooling without the presence of any obvious additional heating, and meanwhile, their volume emission measure (EM) maintains a plateau level for a long time before turning into an evident decay.  Without the need for an additional heating, we propose that the special thermodynamic evolution of the late-phase loops revealed in this flare might arise from loop cross-sectional expansions with height, which are evidenced by both direct measurements from EUV images and by magnetic field extrapolation. By blocking the losses of both heat flux and mass from the corona, such an upward cross-sectional expansion not only elongates the loop cooling time, but also more effectively sustains the loop density, therefore leading to a later-than-expected occurrence of the warm coronal late phase in combination with a sufficiently high late-phase peak. We further verify such a scenario by analytically solving the cooling process of a late-phase loop characterized by a variable cross section.
\end{abstract}


\section{Introduction} \label{sec:intro}

{As one of the most energetic phenomena occurring in the solar atmosphere, solar flares are believed to result from a fast release of energy accumulated in the coronal magnetic fields. Through the so-called magnetic reconnection process \citep{Parker_1963}, the free magnetic energy is rapidly dissipated and consequently goes into} plasma heating, particle acceleration, and bulk mass motion that is usually manifested as a coronal mass ejection \citep[CME; ][]{Lin_2000,Fletcher_2011},  all of which can impose significant space weather disturbances near the Earth \citep{Schwenn_2006}.

According to the standard two-ribbon solar flare model \citep{Carmichael_1964,Sturrock_1966,Hirayama_1974,Kopp_1976}, conventionally known as the CSHKP model, the magnetic reconnection takes place between anti-parallel magnetic field lines stretched out by an erupting magnetic flux rope (MFR), heating post-flare loops and causing brightening flare ribbons at their chromospheric footpoints \citep{Priest_2002,Milligan_2009}. The heating and subsequent cooling of the flare loops meanwhile involve mass circulation between the corona and lower atmospheres \citep{Neupert_1968,Acton_1982,Antiochos_1980,Bradshaw_2010}. In response to these dynamic processes, emissions from the flare loops, as observed in the soft X-ray (SXR) and extreme ultraviolet (EUV) wavebands, typically exhibit an impulsive rise followed by a gradual decay, with the emission peaks occurring sequentially in an order of decreasing formation temperatures of the emissions.

However, the light curves of a real flare event might be much more complex than those predicted by the CSHKP model. Using irradiance observations with the EUV Variability Experiment  \citep[EVE;][]{Woods_2012} on board NASA's \emph{Solar Dynamics Observatory} \citep[\emph{SDO};][]{Pesnell_2012}, \citet{Woods_2011} discovered a second peak of the warm coronal emissions (e.g., the \ion{Fe}{16} 335 {\AA} emission formed at $\sim$ 3 MK) in some solar flares, which occurs several tens of minutes to hours after the flare SXR peak. This phenomenon is hence named as EUV late phase. Around the warm coronal late-phase peak, there is no significant enhancement of the SXR and hot coronal emissions ($\sim$~10 MK). {In spatially resolved observations as taken with the Atmospheric Imaging Assembly \citep[AIA;][]{Lemen_2012} on board \emph{SDO} as well, the late-phase emission is found to originate from another set of higher and longer loops instead of the main flare loops \citep{Woods_2011,Hock_2012,Liu_2013,Li_2014}.}

The occurrence of the EUV late phase implies a special heating history in the late-phase flares. Based on observational case studies, two main mechanisms have been proposed to explain the production of the late phase, namely, long-lasting cooling \citep{Liu_2013,Li_2014,Masson_2017,Dai_2018,Chen_2023} versus additional heating of the late-phase loops \citep{Woods_2011,Hock_2012,Dai_2013,Sun_2013,Zhou_2019,Wang_2020}. In the first scenario, the late-phase loops are heated nearly simultaneously with the main-phase loops, but cool down more gradually due to their significantly longer lengths \citep{Curdt_2004}. This long-lasting cooling naturally results in a delayed and more extended warm coronal emission peak from the late-phase loops. For the latter one, the heating of the late-phase loops takes place well after the main-flare heating. This additional heating should be considerably weak, thus just giving rise to another enhancement of the emissions starting from a medium temperature. By using the zero-dimensional (0D) hydrodynamic model enthalpy-based thermal evolution of loops \citep[EBTEL;][]{Klimchuk_2008,Cargill_2012,Barnes_2016}, the two mechanisms are further numerically validated \citep{Li_2014,Dai&Ding_2018}. It is worth pointing out that both mechanisms may work collectively in a late-phase flare \citep{Sun_2013,Zhong_2021}.

As revealed from the numerical simulations, even with an equal partition of the energy input between the late-phase and main-phase loops, the late-phase peak is still considerably lower than that of the main phase \citep{Dai&Ding_2018}. On the other hand, observations have shown that some late-phase flares do exhibit an extremely large EUV late phase, whose peak is even higher than the main-phase peak \citep{Woods_2011,Liu_2015}. In the frame of the long-lasting cooling scenario, such an extremely large late-phase peak can be attributed to intense heating of the late-phase loops during the flare's main energy release \citep{Dai_2018}. However, the long-lasting cooling seems insufficient to account for a ``later-than-expected" occurrence of the late-phase peak in some events, which raises the necessity for an additional heating, with the heating agent plausibly being an unsuccessfully erupting MFR stopped within the late-phase loop system \citep{Liu_2015,Wang_2016}. 

It should be noted that the conventional estimation of the loop cooling time is based on some simplified assumptions, such as a uniform loop cross-section and classical collision-dominated thermal conduction \citep{Cargill_1995}. In real solar circumstances, a breaking of these conditions may notably influence the loop thermodynamic evolution by altering the energy/mass transport along the loop \citep{Cargill_2022,Reep_2023,Dai_2024}, which makes a longer-than-expected cooling without the need for any additional heating possible. To test this hypothesis, in this paper, we present observations and analysis of a non-eruptive solar flare that exhibits an extremely large EUV late phase. The late-phase loops in this flare are found to be energized by an initial strong heating due to the failed eruption of an MFR, and their apparent longer-than-expected cooling is more likely to arise from a cross-sectional expansion with height rather than additional heating. The rest of the paper is organized as follows. In Section \ref{sec:def} we identify the extremely large late phase. The heating and cooling processes of the late-phase loops are explored in Sections \ref{sec:heat} and \ref{sec:cool}, respectively. Finally, we discuss the results and draw our conclusions in Section \ref{sec:dis}.

\section{Extremely Large EUV Late Phase} \label{sec:def}
In this study, we focus on an X1.8-class solar flare occurring on 2012 October 23 from NOAA active region (AR) 11598. For its complex magnetic configuration and profound impacts on various atmospheric layers, the flare has been extensively studied from a variety of aspects in the literature \citep[e.g.,][]{Yang_2015,Sharykin_2017,Watanabe_2020,Wu_2023,Liu_2024}. 

\begin{figure}
\epsscale{0.9}
\plotone{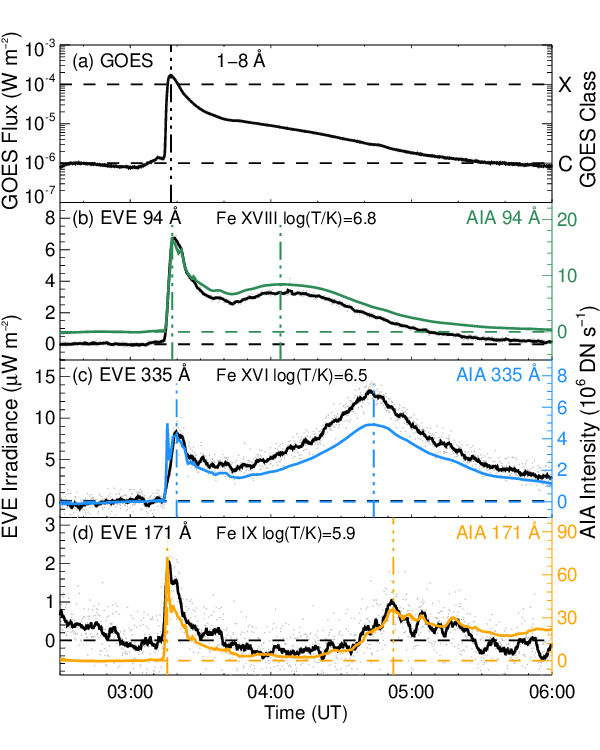} 
\caption{Time profiles of the \emph{GOES} 1--8 {\AA} SXR flux (a), and the background-subtracted EVE full disk irradiance (black) and AIA sub-region intensities (colored) in several lines and passbands (b)--(d) for the 2012 October 23 X1.8 flare. The vertical dashed-dotted-dotted lines mark conspicuous peaks of the \emph{GOES} flux and AIA intensities, and the horizontal dashed lines in panels (b)--(d) denote the background levels for EVE and AIA data.  Note that a 100--150 s smoothing boxcar is applied to the EVE raw data points (dot signs), and the AIA sub-region covers a field-of-view (FOV) of $210\arcsec \times 210\arcsec$ enclosing the flare-hosting AR.}
\label{fig:eve}
\end{figure}

Figure \ref{fig:eve}(a) displays the time profile of the \emph{Geostationary Operational Environmental Satellite} (\emph{GOES}) 1--8 {\AA} SXR flux for this event. The flare begins at around  03:13 UT, promptly reaches its peak at 03:17:23 UT, and then turns into a gradual decay lasting for nearly three hours. It is noted that the decrease of the SXR flux exhibits a dual-decay pattern, with the slope of the light curve experiencing an obvious flattening at around 03:30 UT.

To investigate the flare emissions at other temperatures, in Figures \ref{fig:eve}(b)--(d) we plot the background-subtracted EVE irradiance for three spectral lines. EVE measures Sun-as-a-star spectra from 1 {\AA} to 1050 {\AA} with 1 {\AA} spectral resolution and 10 s time cadence, from which the line irradiance data are derived by spectral integration over specified spectral windows. As shown in the figure, first, the cool coronal \ion{Fe}{9} 171 {\AA}  ($\log(T/\mathrm{K})\sim5.9$) line shows an impulsive rise and peaks before the \emph{GOES} SXR maximum, reflecting a rapid response of the lower atmosphere to impulsive flare heating. Second, the moderately hot coronal \ion{Fe}{18} 94 {\AA} ($\log(T/\mathrm{K})\sim6.8$) and warm coronal \ion{Fe}{16} 335 {\AA} ($\log(T/\mathrm{K})\sim6.5$) lines peak sequentially shortly after the SXR peak, indicating a relatively fast cooling process of flare loops. These patterns are common to typical solar flares. Third and differently, in all coronal lines, there are another prominent emission peak well after the corresponding main flare peak. For EVE 335~{\AA}, the time lag and emission ratio between the two peaks are $\sim$83 minutes and $1.75\pm0.40$, respectively, conforming to the criterium for an extremely large EUV late phase \citep{Liu_2015}. In the hotter (cooler) lines, the second peak appears earlier (later), which may reflect another cooling process in late-phase loops. 

\begin{figure*}
\epsscale{0.85}
\plotone{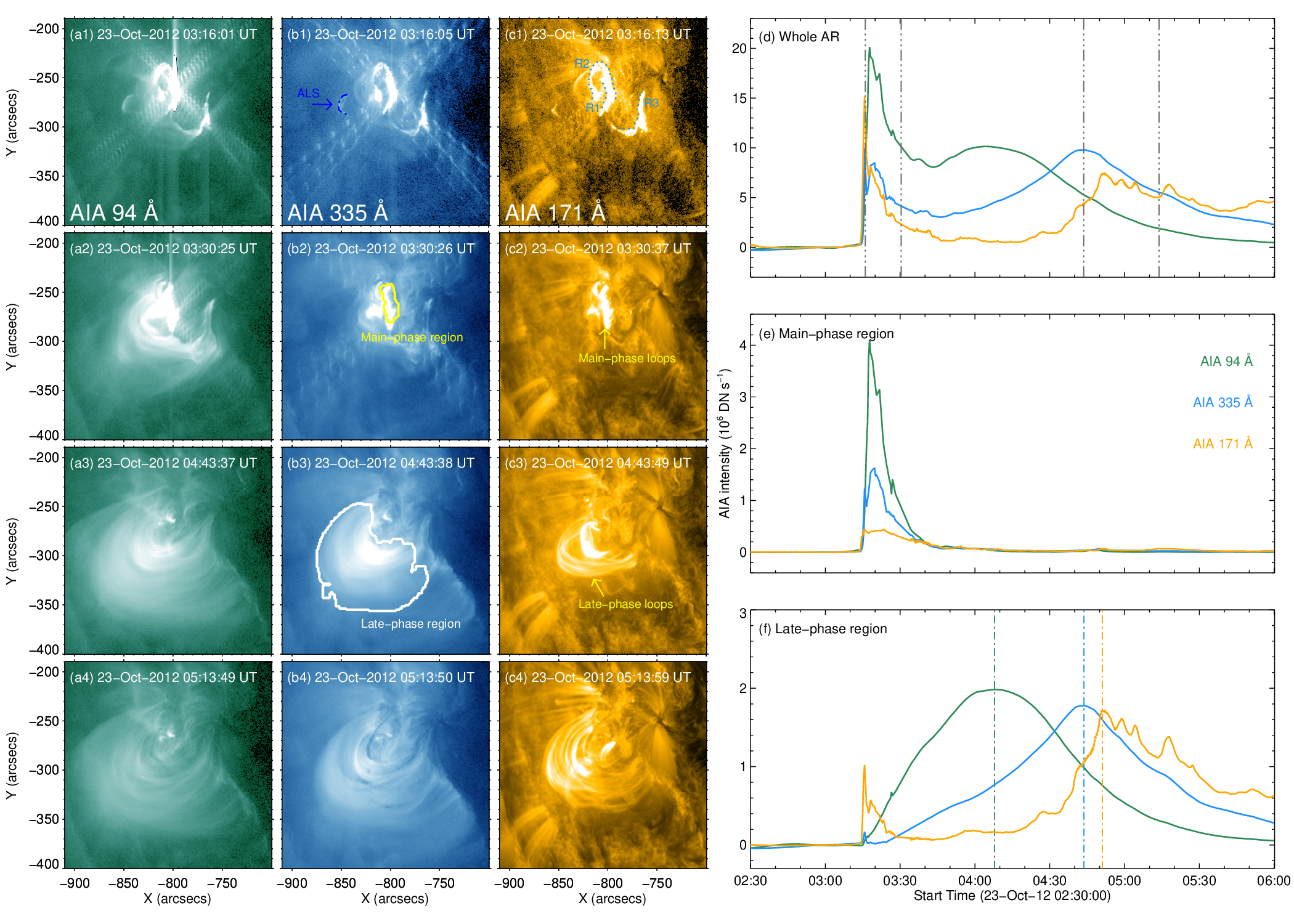}    
\caption{Evolution of the flare observed by AIA. Left: snapshots of the flare evolution in AIA 94 (a1)--(a4), 335 (b1)--(b4), and 171 {\AA} (c1)--(c4), respectively. The yellow and white contours in panels (b2) and (b3) enclose the main-phase and late-phase regions, respectively. Some other characteristic structures are also traced and highlighted; see the text for a detailed description. Right: light curves of the three AIA passbands over the whole AR (d), the main-phase region (e), and the late-phase region (f), respectively. The vertical dashed-dotted-dotted lines in panel (d) mark the four moments around which the left column snapshots are displayed (each row for one moment), and the vertical dashed-dotted lines in panel (f) outline the peaks of the AIA profiles for the late-phase region. {An animation of this figure is available. The animation begins at approximately 02:30:07 UT and ends at 06:00:07 UT on 2012 October 23. The real-time duration of the animation is 21 s.}} 
\label{fig:elp}
\end{figure*}

To consolidate the identification of the extremely large EUV late phase, we then resort to AIA imaging observations. AIA provides full-disk images of the transition region (TR) and corona in ten passbands with a pixel scale of 0.6{\arcsec} and a cadence of 12 or 24 s. Figure \ref{fig:elp}  demonstrates the evolution of the flare in the AIA passbands of 94, 335, and 171 {\AA}, which have a one-to-one correspondence with the EVE lines plotted in Figure \ref{fig:eve}. The flare event is driven by the eruption of an arcade-like structure (ALS, pointed out in Figure \ref{fig:elp}(b1)) from $\sim$03:14 UT. By inspecting pre-event AIA images in 304 {\AA} (not shown here), the predecessor of the ALS is identified as a filament, which, according to the magnetic modeling by \citet{Yang_2015}, corresponds to an MFR embedded in a fan-dome structure. Under such complex magnetic configuration, multiple magnetic reconnections are invoked, producing three main flare ribbons  (outlined by the dotted lines and labeled as R1--R3 in Figure \ref{fig:elp}(c1)), with two elongated ribbons inside (R1) and outside (R3) a circular one (R2). In AIA, the inner ribbon is barely distinguishable from the circular ribbon, but can be clearly discerned in \ion{Ca}{2} H emission \citep{Yang_2015}. Connecting these flare ribbons, multiple sets of flare loops are observed in the corona. Between the inner ribbon and west part of the circular ribbon, where the filament-hosting MFR is initially located, compact flare loops first brighten up (pointed out by the arrow in Figure \ref{fig:elp}(c2)).  With the ascent of the erupting MFR, flare loops then appear over a much larger extent (indicated by the arrow in Figure \ref{fig:elp}(c3)), which predominately link the east part the circular flare ribbon to the remote ribbon. Owing to their distinct lengths, the two sets of flare loops should belong to main-phase loops and late-phase loops, respectively. In hot AIA 94 {\AA}, the appearances of the main-phase loops and late-phase loops overlap in time, while in the cooler passbands (especially in AIA 171 {\AA}), they are temporally separated.  Note that the re-appearance of the late-phase loops very late in AIA 94~{\AA} is more likely to result from a secondary response of this passband to \ion{Fe}{10} ions ($\log(T/\mathrm{K})\sim6.0$). 

Besides the morphological evolution, we further trace light curves over some regions of interest with AIA\@. Figure \ref{fig:elp}(d) plots the intensity profiles of the whole AR in the three AIA passbands, which are calculated by summing the intensities of all pixels within a region of $x \in [-910\arcsec, -700\arcsec]$ and $y \in [-400\arcsec, -190\arcsec]$. For comparison, the AIA intensity profiles are also individually over-plotted in Figures \ref{fig:eve}(b)--(d), which show a close similarity to the corresponding EVE light curves. In particular, the AIA 335 {\AA} profile also reveals an extremely large late-phase peak, with the peak time nearly simultaneous with that in EVE 335 {\AA}. Such consistency means that the EUV variabilities detected by EVE should predominately originate from the AR\@.

Based on the AR light curve in AIA 335 {\AA}, we use the method proposed by \citet{Chen_2020} to identify and extract the main-phase and late-phase regions. First, we pick up two AIA 335 {\AA} images taken at the main-phase peak and late-phase peak, respectively. Then, we subtract the latter image from the former one. In the different image, the main-phase (late-phase) region is labeled as pixels with values higher (lower) than the average value of all positive (negative) pixels. Last, we apply a morphological open operation to filter out patchy pixels from both regions. The resultant main-phase and late-phase regions are enclosed by the yellow contour in Figure \ref{fig:elp}(b2) and the white contour in Figure \ref{fig:elp}(b3), respectively. In comparison, the late-phase region occupies a projection area of $4.28\times10^{19}$ cm$^2$, much larger than that of the main-phase region ($2.36\times10^{18}$ cm$^2$). To test the robustness of our region extraction, we individually calculate the AIA intensity profiles in these two subregions. As shown in Figures \ref{fig:elp}(e) and (f), the emissions from the identified main-phase region well characterize the main-phase peak, as do those from the identified late-phase region for the late-phase peak. 

\section{Failed MFR Eruption and Heating of the Late-phase Loops} \label{sec:heat}
According to the location of the main-phase and late-phase regions, magnetic reconnections should take place both below and above the erupting MFR, heating the main-phase loops and late-phase loops, respectively. The main-phase reconnection is a typical standard-flare reconnection, as inferred from the morphology of the main-phase loops as well as the elongation of the flare ribbons. As to the late-phase reconnection, it is supposed to happen within a large-scale dome-structured quasi-separatrix layer (QSL) squeezed by the ascending MFR \citep[cf.][]{Yang_2015}. In this sense, the way the MFR interacts with the overlying QSL may remarkably affect the heating strength of the late-phase loops.

\begin{figure*}
\epsscale{0.8}
\plotone{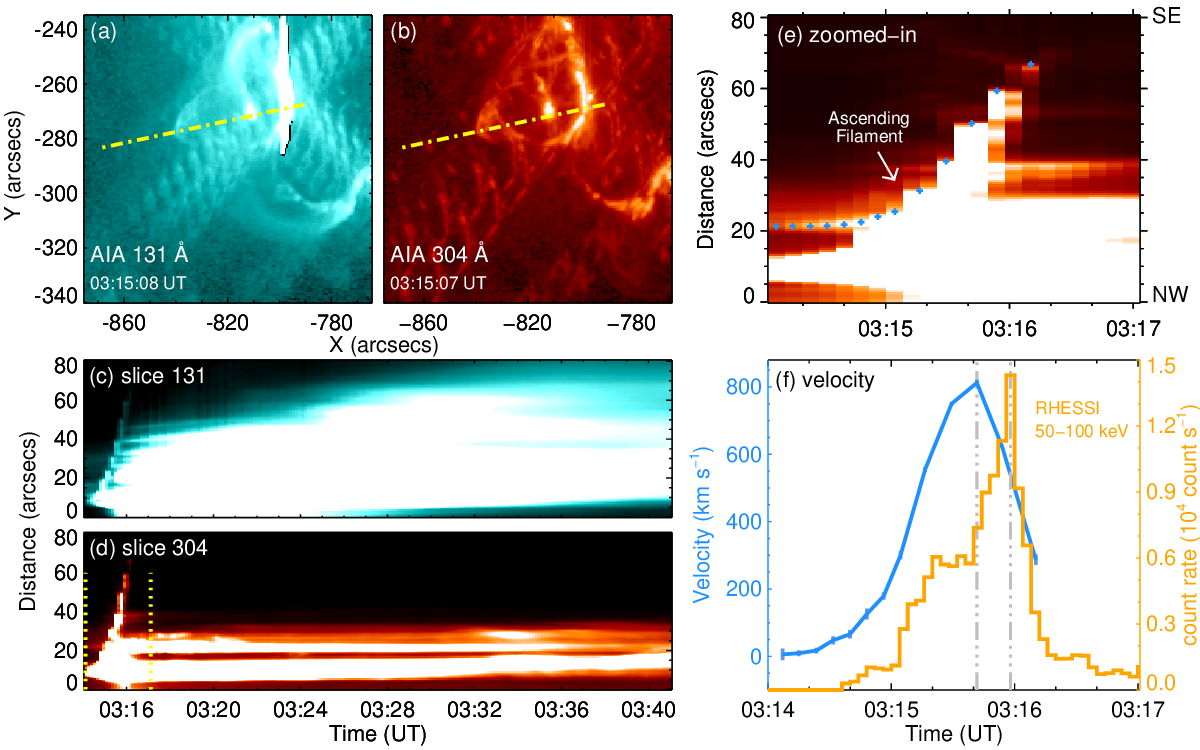}
\caption{Eruption process of the filament-hosting MFR. Panels (a) and (b) show the snapshots of the MFR eruption in AIA 131 and 304 {\AA}, respectively. Panels (c) and (d) display the time--distance stack plots of the AIA 131 and 304 {\AA} intensities along the slice (dashed-dotted line) drawn in panels (a) and (b). A zoomed-in view of the part bounded by the dotted lines in panel (d) is also displayed in panel (e), in which the position of the MFR apex is traced by the plus signs. Panel (f) plots the variation of the MFR velocity (in blue) according to the time--distance measurements shown in panel (e), as well as the \emph{RHESSI} 50--100 keV count rate for this flare (in orange). The vertical dashed-dotted-dotted lines mark the peak times of the two profiles.}
\label{fig:time}
\end{figure*}

To this end, we select a slice passing across the apex of the erupting MFR (indicated by the yellow dashed-dotted line in Figures \ref{fig:time}(a) and \ref{fig:time}(b)), and trace the evolution of AIA intensities along this slice. The time--distance stack plots for the AIA 131 {\AA} (\ion{Fe}{21}, $\log(T/\mathrm{K})\sim7.1$) and 304 {\AA} (\ion{He}{2}, $\log(T/\mathrm{K})\sim4.7$) passbands are plotted in Figures \ref{fig:time}(c) and \ref{fig:time}(d), respectively. Both plots reveal an evident ascending motion of the MFR, which is manifested as a curved stripe brightening up during an interval of 03:14--03:16 UT\@. The co-appearance and nearly the same trajectory of the MFR apex in both AIA passbands suggest a multi-thermal nature of the MFR system \citep{Wang_2022}. Following the passage of the MFR, quasi-stationary brightening structures are observed in AIA 131 {\AA} (Figure \ref{fig:time}(c)). Since they are invisible in AIA 304 {\AA} (the persistent brightening stripes in Figure \ref{fig:time}(d) are mainly flare ribbons), the brightening structures obviously reflect the production of high temperature late-phase loops, hence corroborating the causality between the eruption of the MFR and the heating of the late-phase loops.

To quantify the kinematic evolution of the ascending MFR, we track the displacement of the MFR apex by visual inspection (with the results over-plotted as the plus signs in a zoomed-in view of the AIA 304 {\AA} time--distance plot in Figure \ref{fig:time}(e)). {To minimize the impact of subjectivity inherent in manual inspection, we repeat the position measurement ten times at each point, by which the average of the measurements is taken as the measured value, and so does the corresponding standard deviation as the uncertainty.}  Based on the time--distance data, we calculate the velocity of the MFR using a numerical differentiation with three-point Lagrangian interpolation. As shown in Figure \ref{fig:time}(f), the MFR first experiences an impulsive acceleration. Within a short period from 03:14:07 to 03:15:42 UT, its velocity quickly increases from nearly zero to approximately 800 km s$^{-1}$. Afterward, the MFR turns into an even stronger deceleration before it finally disappears from the selected slice (after $\sim$03:16:10 UT).

In Figure \ref{fig:time}(f), we also plot the 50--100 keV hard X-ray (HXR) count rate obtained with the \emph{Reuven Ramaty High Energy Solar Spectroscopic Imager} \citep[\emph{RHESSI};][]{Lin_2002}, which is believed to characterize the strength of the main-phase reconnection. It is seen that the increase of the HXR count rate basically synchronizes with the velocity evolution of the MFR, except that its peak is delayed by $\sim$ 20 s with respect to the velocity peak, a typical feature revealed in non-eruptive flares \citep{Huang_2020}. The initial synchronicity between the MFR velocity and the HXR count rate should imply a positive feedback between the MFR acceleration and the main-phase connection \citep{Zhang_2001}. Nevertheless, the earlier occurrence of the velocity peak indicates that a strong confinement from the above has started to impose on the MFR while the main-phase reconnection is still developing.  In this case, the extremely large MFR deceleration before the HXR peak ($\sim-15$ km s$^{-2}$, nearly 55 times greater than the solar gravitational constant) suggests an extremely large drag force acting on the MFR, or in other words, an extremely large squeeze force the MFR acts on the overlying QSL, which would greatly enhance the current inside it, consequently leading to an intense late-phase reconnection/heating \citep{Dai_2018}. 

Since the velocity of the MFR does not decrease to zero before its disappearance from the slice (see Figures \ref{fig:time}(e) and (f)), one would conjecture that the MFR could be deflected to another direction. Nevertheless, by inspecting the full AIA images, we find no evidence of the MFR deflection. Instead, the erupting MFR is finally merged into the late-phase loop system. Meanwhile, no CMEs are observed in the outer corona. All these facts suggest a failed MFR eruption. During this process, the MFR itself can also reconnect directly with the overlying field lines. Such an external reconnection not only erodes the MFR, but also makes an additional contribution to the late-phase heating \citep{Chen_2023a}.

\section{Longer-than-expected Cooling of the Late-phase Loops} \label{sec:cool}
\subsection{Cooling of the Overall Late-phase Region} \label{sec:time}
Using observations with the six AIA coronal passbands, we conduct differential emission measure (DEM) analysis to the AR\@. As a pre-processing to enhance the signal-to-noise ratio, we first spatially re-bin the pixels of each AIA image by a factor of 2, and take the time average of every five images as an input image.  Then we adopt the sparse algorithm \citep{Cheung_2015,Su_2018} to perform the DEM inversion over a temperature range of $\log(T/\mathrm{K}) \in[5.5, 7.6]$, with the grid spacing set to be 0.05dex. {The validity of the AIA-based DEM inversion is later verified by cross-checking the inversion results with EVE observations  (see Appendix \ref{appA}).}

\begin{figure*}
\epsscale{0.8}
\plotone{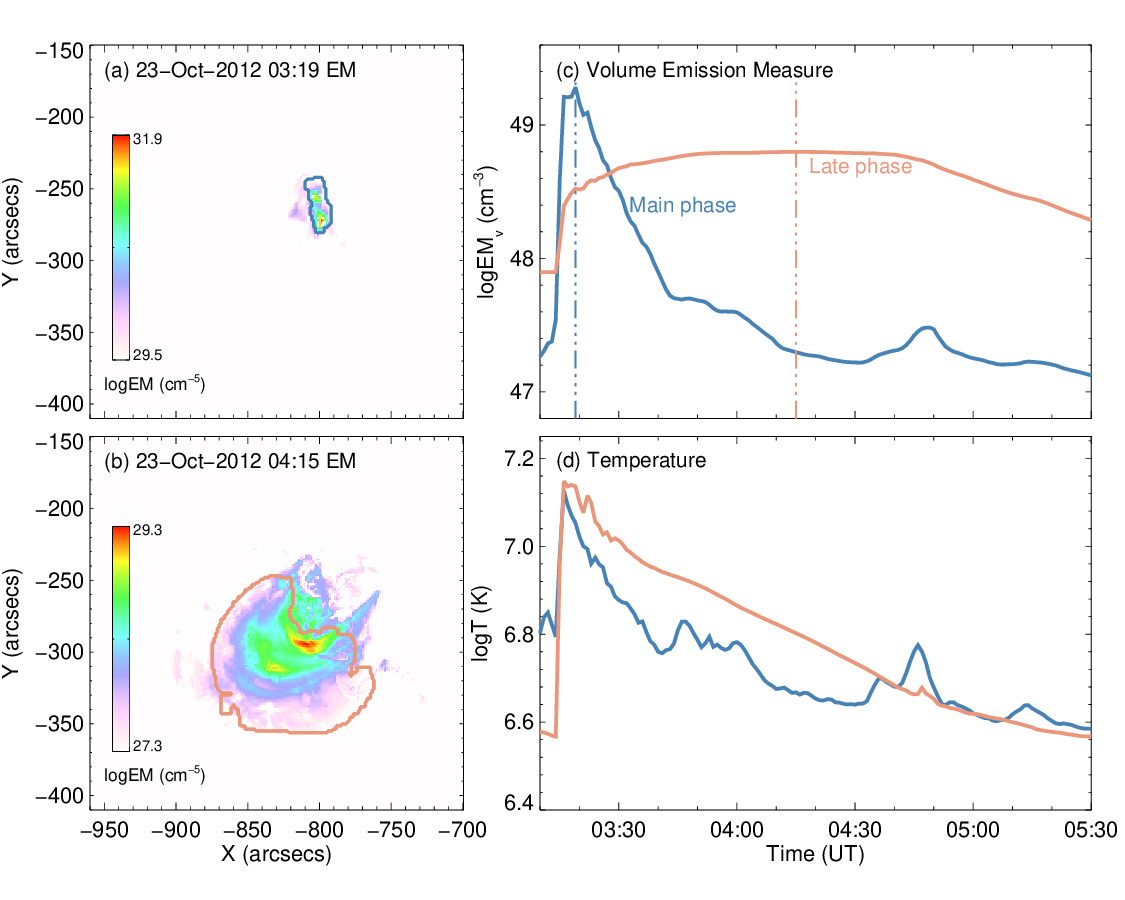}
\caption{DEM inversion results for the flare. Left: EM maps of the AR at two moments when the main-phase (enclosed by the blue contour in panel (a)) and late-phase (enclosed by the orange contour in panel (b)) regions are well developed, respectively. Right: temporal evolutions of the volume EM (c) and DEM-weighted temperature (d) for the two regions (discriminated by the colors). The vertical dashed-dotted-dotted lines in panel (c) indicate the times of maximum volume EM for the main-phase and late-phase regions, respectively.}
\label{fig:em}
\end{figure*}

Based on the inversion results, we can obtain the emission measure (EM) for each pixel by
\begin{equation}
EM=\int DEM(T)\,dT,
\end{equation}
as well as the volume EM ($EM_V$) and DEM-weighted temperature ($\bar{T}$)  over a specified region by
\begin{equation}
EM_V=\sum_iS_{\mathrm{pix}}EM_i=S_{\mathrm{pix}}\int \sum_iDEM_i(T)\,dT
\end{equation}
and
\begin{equation}
\bar{T}=\frac{\int \sum_iDEM_i(T)T\,dT}{\int \sum_iDEM_i(T)\,dT},
\end{equation} 
where the subscript $i$ denotes the $i$th pixel in the region of interest, and $S_{\mathrm{pix}}$ is the area of the pixel (the same for all pixels). Figures \ref{fig:em}(a) and (b) display the EM maps at two moments when the main-phase and late-phase regions are well developed, respectively. It is seen that the emitting materials are predominately localized in our identified main-phase/late-phase region, further verifying the validity of our region identification and extraction algorithm. To quantitatively trace the thermodynamics of the two regions, in Figures \ref{fig:em}(c) and (d) we plot the temporal evolutions of their volume EM and DEM-weighted temperature. Compared with the fluctuating decay of the temperature in the main-phase region, the late-phase region shows a rather smooth and monotonic cooling, whose duration extends more than two hours. Meanwhile, its volume EM maintains a plateau level for over one hour before turning into an evident decay, also significantly longer than the duration of EM enhancement in the main-phase region.

By approximating the late-phase region with a set of (identical) characteristic loops, we can theoretically estimate its overall cooling time. {Assuming that conductive losses dominate first and radiation takes over later on, \citet{Cargill_1995} investigated the cooling process of a flare loop, and presented a ready-to-use formula to estimate the overall cooling time of the loop. Here, we extend the previous analytical studies on different stages of the loop cooling (see Appendices \ref{appB1} and \ref{appB2}), and improve the Cargill's loop cooling time formula (see Appendix \ref{appB3}), which is now modified as}
\begin{equation}
\tau_{\mathrm{cool}} = 2.04\times 10^{-2}L^{5/6}T_{0}^{-1/6}n_{0}^{-1/6}\ \ \mathrm{[s]},\label{tcool_eqn}
\end{equation}
where $L$ is the half-length of the characteristic late-phase loop, and $T_0$ ($n_0$) is the loop temperature (density) at the start of the cooling. As a rough estimation,  $L=S^{1/2}$, where $S$ is the projection area of the late-phase region, $T_0$ is approximated by the maximum DEM-weighted temperature of the region, and $n_0$ is calculated from corresponding volume EM by $(EM_V/V)^{1/2}$, where $V$ is the volume of the late-phase region given as $V=S^{3/2}$. For the late-phase region identified in this case, $L=65$ Mm, $T_0=14$ MK, and $n_0=3.0\times10^9$ cm$^{-3}$, which results in a theoretically estimated cooling time of 87 minutes. By comparison, the observed cooling time of the late-phase region is significantly longer, indicating a longer-than-expected cooling.

\subsection{Cooling of a Representative Late-phase Loop and Evidence of Cross-sectional Expansion}
\begin{figure*}
\epsscale{0.8}
\plotone{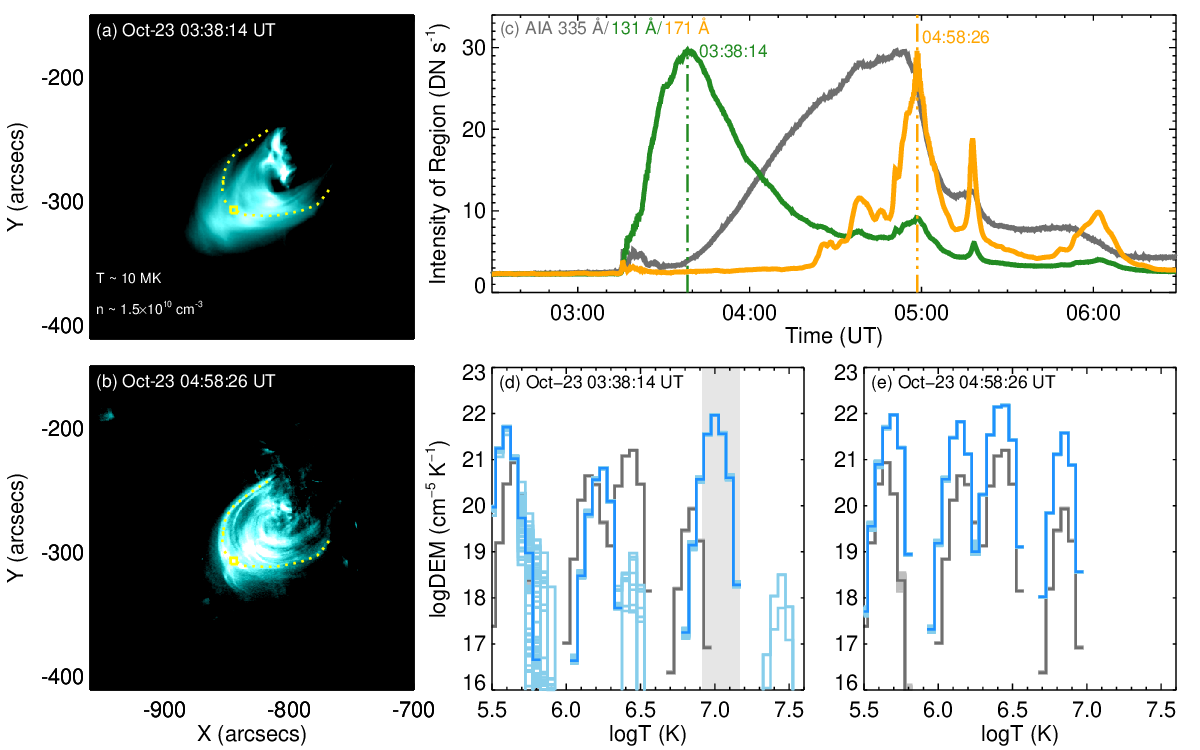} 
\caption{Evolution of the representative late-phase loop. Left: snapshots of the AIA 131 {\AA} images showing the late-phase region, where the representative late-phase loop is traced by the dotted line and overlaid by a $5\arcsec\times5\arcsec$ box near its apex. Right: thermodynamic evolution for the selected box. Panel (c) plots the AIA 131, 335, and 171 {\AA} light curves over the box, where the vertical dashed-dotted-dotted lines represent the peak times in AIA 131 and 171 {\AA}, respectively. Panels (d) and (e) depict the corresponding DEM distributions (plus the results of 100 Monte Carlo simulations, in blue) at these two peak times. For comparison, the pre-event DEM for the box (in gray) is also over-plotted.}
\label{fig:cool}
\end{figure*}

\begin{figure*}
\epsscale{0.85}
\plotone{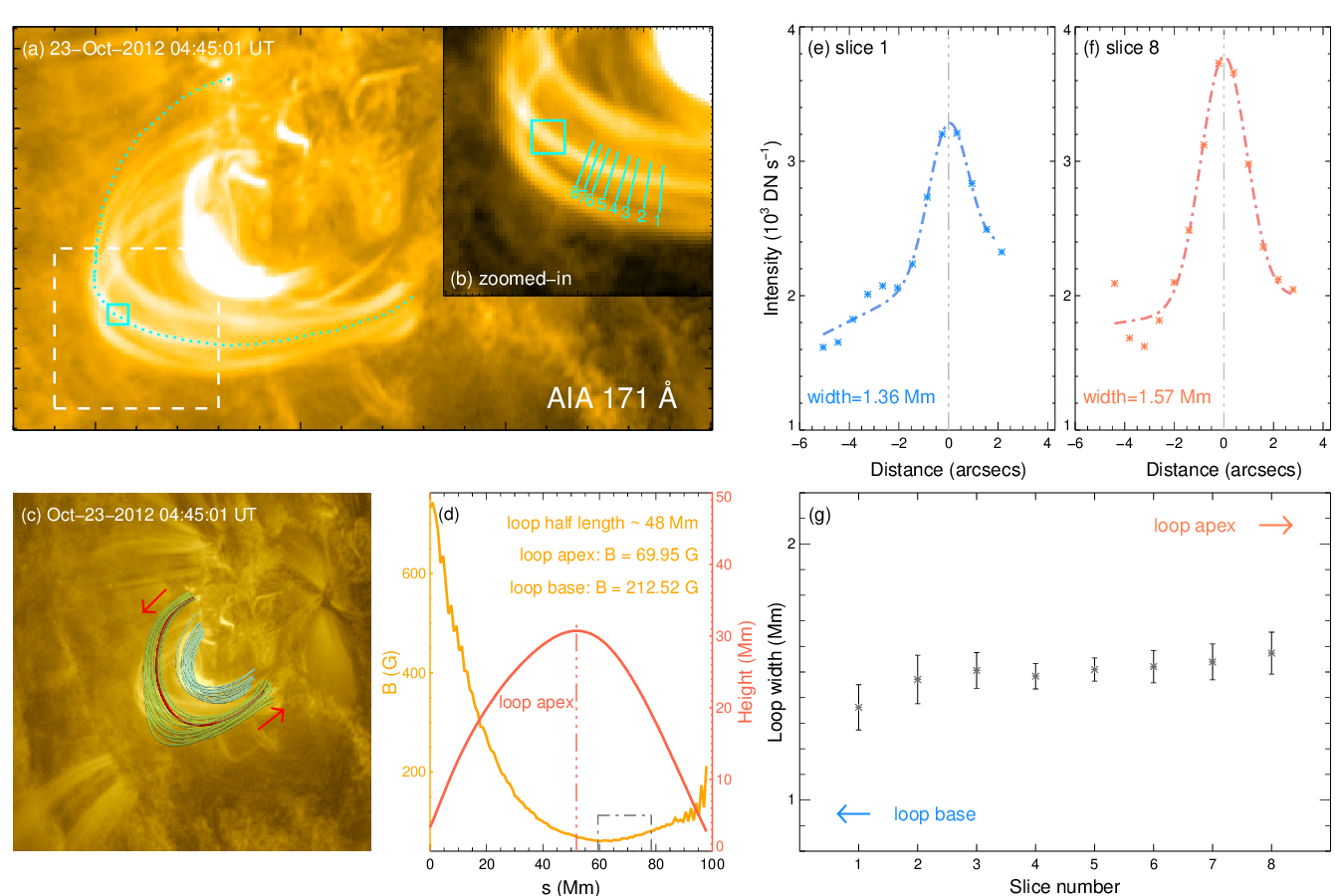}
\caption{Magnetic modeling and width measurements of the representative late-phase loop. Panel (a) shows an AIA 171 {\AA} image of the late-phase region, with the dotted line tracing the representative late-phase loop. A zoomed-in view of the region enclosed by the dashed box is displayed in panel (b), where we put a total of eight slices (numbered from 1 to 8 toward the close-to-apex box) locally perpendicular to the loop spine for width measurement. Panel (c) overlays some magnetic field lines traced from the NLFFF extrapolation, and panel (d) plots the variations of the magnetic strength and height along a proxy field line (the red line in panel (c)) that best matches the representative late-phase loop. The dashed-dotted box in panel (d) represents the extent of the loop segment used for width measurement. Panels (e) and (f) demonstrate the results of Gaussian fitting (dashed-dotted lines) to the AIA intensities (asterisks) along Slices 1 and 8, respectively, with the fitted loop widths labeled in the legend as well, and panel (g) shows the tendency of loop width variation based on the fitting results for all slices. Note that the error bars are estimated from the fitting uncertainties.}
 \label{fig:expansion}
\end{figure*}

To consolidate the claim of the longer-than-expected cooling pattern, we further investigate the cooling of a representative late-phase loop that permits more reliable measurements of the loop properties. The loop is picked up by visual inspection. In Figures \ref{fig:cool}(a) and (b) (as well as in Figure \ref{fig:expansion}(a)) we trace the spine of the loop with a dotted line, and overlay a $5\arcsec\times5\arcsec$ box on the close-to-apex part of the loop, whose evolution serves as a proxy for the whole loop evolution. 

The light curves of AIA 131, 335, and 171 {\AA} over the selected box are plotted in Figure \ref{fig:cool}(c). It is seen that peak of the AIA 171 {\AA} profile (04:58 UT) is delayed from that of AIA 131 {\AA} (03:38 UT) by 80 minutes. The corresponding DEM distributions at these two peak times are displayed (in blue) in Figures \ref{fig:cool}(d) and (e), respectively, and as a reference, we also over-plot the pre-event DEM for this box (in gray). Compared with the pre-event DEM, the DEM at the AIA 131 {\AA} peak exhibits a prominent bump around $\log(T/\mathrm{K})=7.0$ (highlighted by the shaded region in Figure \ref{fig:cool}(d)), indicating a heating of loop plasma to high temperatures. At the AIA 171 {\AA} peak, nevertheless, this high temperature bump totally disappears, and discernible DEM enhancements have shifted down to lower temperatures (Figure \ref{fig:cool}(e)). During this period, the AIA 335 {\AA} light curve largely shows a gradual rise followed by fast decay (Figure \ref{fig:cool}(c)). In this sense, a cooling process of the late-phase loop is convincingly established, and we tactically take the time delay between the two peaks, 80 minutes, as the observed loop cooling time \citep[cf.][]{Dai&Ding_2018,Chen_2023}.   

We use Equation (\ref{tcool_eqn}) once again to theoretically estimate the cooling time of the representative late-phase loop, but here the loop properties are evaluated in a more reasonable way. For consistency, we take the AIA 131 {\AA} peak time as the start time of the loop cooling, at which the contribution of heated loop plasma to the DEM comes predominately from the high temperature bump. By isolating this component, we obtain a column EM of $3.4\times10^{28}$ cm$^{-5}$ as well as a DEM-weighted loop temperature of 10~MK (taken as the initial temperature). With the obtained EM, the loop density is now determined by
\begin{equation}
 n_e=\sqrt{\frac{EM}{d}},
\end{equation}
where $d$ is the thickness of the loop along the line-of-sight (LOS), which is also equivalent to the loop width against the plane of the sky. Following the method of \citet{Mandal_2024}, we apply a Gaussian fitting to the intensities perpendicular to the loop spine, and take the full width at half maximum (FWHM) of the resultant Gaussian function as the loop width. Near the loop apex, the loop width is found to be 1.6 Mm (see Figure \ref{fig:expansion}(f)), which yields an initial loop density of $1.5\times10^{10}$~cm$^{-3}$.

As to the determination of the loop half-length, we rely on nonlinear force-free-field (NLFFF) extrapolation rather than visual inspection, whose accuracy is usually affected by the projection effect. Here, we choose a photospheric vector magnetogram obtained at 04:48 UT with the Helioseismic and Magnetic Imager \citep[HMI;][]{Scherrer_2012,Schou_2012} also on board \emph{SDO} as the bottom boundary. After a remapping of the original magnetogram in a cylindrical equal-area (CEA) projection \citep{Gary_1990} followed by a pre-processing to minimize the magnetic force and torque on the bottom boundary \citep{Wiegelmann_2006}, we use the magneto-frictional method \citep{Guo_2016} to perform the NLFFF extrapolation. In Figure \ref{fig:expansion}(c) we plot some magnetic field lines traced from the extrapolation result, which show a rather good match with the late-phase loops seen in AIA\@. Based on the degree of match, we highlight one field line (red line in Figure \ref{fig:expansion}(c)) to mimic the representative late-phase loop selected above. Tracing this field line in the three-dimensional (3D) computation domain yields a full-length of 98~Mm, as well as an apex position located near its midpoint (Figure \ref{fig:expansion}(d)). In this way, the half-length of the representative late-phase loop is reliably estimated to be 49~Mm.

With the above loop parameters, the theoretical cooling time of the representative late-phase loop turns out to be 55 minutes. As a quantitative comparison, the observed cooling time ({80 minutes}) is longer than the theoretical value by 45\%. {To test whether such a discrepancy is of physical significance or not, we further carry out an error analysis of the loop parameters in determining the cooling time. Based on 100 Monte Carlo simulations of the DEM inversion, it is seen that the high temperature DEM bump at the start of the cooling is excellently constrained (see Figure \ref{fig:cool}(d)), from which the derived initial loop temperature and density are found to vary by only a few percent. Considering the extremely weak dependence of the cooling time on temperature and density (both with a power index of -1/6), the uncertainty of the cooling time estimation should solely come from the loop length estimation (in a power dependence of 5/6). As shown in Figure \ref{fig:expansion}(c), besides the proxy magnetic field line that mimics the representative late-phase loop, we have traced another 15 field lines from a random start point in the vicinity of its footpoint. The half-lengths of these field lines are found to range from 46 to 68 Mm, corresponding to theoretical cooling times from 53 to 73 minutes, the maximum of which is still $\sim$10\% shorter than the observed value. On the other hand, to account for an observed cooling time of 80 minutes, the half-length of the loop needs to be 84~Mm, $\sim$15\% longer than the maximum value among these randomly traced field lines. In this sense, a longer-than-expected cooling of the representative late-phase loop is physically consolidated.} 

Finally, we conduct an analysis of the variation of loop width along the representative late-phase loop. We pick up a segment of the loop that has sufficient intensity and meanwhile is less affected by the contamination from other loops. Along this segment, we put a total of eight slices locally perpendicular to the loop spine (Figure \ref{fig:expansion}(b)), and use the above-mentioned method to measure the loop width for each slice. As shown in Figure \ref{fig:expansion}(g), the loop width shows a clear tendency of increasing with slice number, indicating an expansion of the loop cross section with height. From Slice 1 to Slice 8, the loop width increases by 15\% (see Figures \ref{fig:expansion}(e) and (f)), which is equivalent to a cross-sectional expansion of 33\% over the selected segment. 

Assuming a conservation of magnetic flux along the loop, the expansion of the cross section should reflect a decay of the loop-aligned magnetic field with height. In Figure \ref{fig:expansion}(d) we also plot the variation of magnetic strength along the proxy field line for the late-phase loop. Through visual match, we locate the position of the selected loop segment with a dashed-dotted box drawn in the figure. Over this part, the magnetic strength decreases by 25\%, which is converted to a 33\% increase of the cross section, in good agreement with the result derived from the direct measurements. In passing, we note that the selected segment is located in the weak half part of the loop in terms of the magnetic strength, with a base-to-apex ratio of 3 versus the value of over 10 for the strong half part.  Even so, the loop cross section can still expand by a factor of 3 over this side, possibly giving rise to a non-negligible influence on the thermodynamic evolution of the loop. 
 
\section{Discussion and Conclusions} \label{sec:dis}
Using observations mainly with \emph{SDO}, we analyze an X1.8-class solar flare occurring on 2012 October 23 from a famous AR 11598. The flare exhibits an extremely large late phase in the warm coronal EUV emissions, with the peak intensity more than 1.4 times that of main flare peak (see Figure \ref{fig:eve}(c)). Flare loops from the late-phase region brighten up sequentially in AIA passbands of decreasing temperatures (see Figure \ref{fig:elp}), and a causality between the heating of the late-phase loops and the eruption of an MFR is established (see Figure \ref{fig:time}). These features favor the long-lasting cooling scenario for the production of EUV late phase from an early heating \citep{Liu_2013,Li_2014,Masson_2017,Dai_2018,Chen_2023}, with the heating agent presumably being an MFR.

The ascending MFR first experiences a very impulsive acceleration, nearly synchronous with the increase of the HXR count rate (see Figure \ref{fig:time}(f)). Such a synchronicity suggests a positive feedback between the MFR acceleration and the main-phase magnetic connection \citep{Zhang_2001}, since the main-phase reconnection takes place between the magnetic field lines stretched out by the erupting MFR, and in turn adds poloidal flux to the MFR, further facilitating its ascent. Embedded in a fan-dome structure, the ascending MFR also pushes and squeezes the above dome-structured QSL, causing large-scale magnetic reconnections that heat the late-phase loops. Such an interaction between the MFR and overlying QSL naturally imposes a drag force on the MFR. It is found that after the impulsive acceleration, the erupting MFR quickly turns to an even stronger deceleration. This implies an extremely large drag force acting on the MFR, or in other words, an extremely large squeeze force the MFR acts on the overlying QSL, which would greatly enhance the current inside it, consequently leading to an intense late-phase reconnection/heating \citep{Dai_2018}. On the other hand, the continuous MFR deceleration meanwhile suppresses the main-phase reconnection (since the HXR peak is just delayed from the MFR velocity peak by 20 s), hence biasing the energy partitioning more toward the late-phase heating. Moreover, the MFR fails to escape and is finally merged into the late-phase loop system. During this process, the MFR itself can also reconnect directly with the overlying field lines. Such an external reconnection not only erodes the MFR, but also makes an additional contribution to the late-phase heating \citep{Chen_2023a}. All these factors may account for the production of the extremely large late-phase peak seen in this event, and are consistent with the statistical result that a prominent EUV late phase is more likely to appear in non-eruptive flares \citep{Wang_2016}.

Based on the DEM analysis, it is found that the late-phase loops experience an extended cooling, whose duration is significantly longer than the theoretically estimated cooling time. In the past, such a longer-than-expected cooling was conventionally attributed to an additional heating taking effect during the decay phase \citep[e.g.,][]{Ryan_2013,Brose_2022}. In this case, however, the DEM-weighted temperature of the late-phase region shows a rather smooth and monotonic decay (see Figure \ref{fig:em}(d)), to a large extent ruling out the presence of an obvious heating during this stage. It should be noted that the conventional estimation of the loop cooling time relies on some simplified assumptions, e.g., a uniform loop cross-section, which is not necessarily satisfied in real coronal loops. Here, for a representative late-phase loop that more credibly exhibits the longer-than-expected cooling pattern, both the direct measurements and the NLFFF extrapolation give consistent evidence of a loop cross-sectional expansion with height (see Figure \ref{fig:expansion}).  Without the need for an additional heating, we propose that the special thermodynamic evolution of the late-phase loops revealed in this flare might arise from loop cross-sectional expansions. 

According to the loop cooling theory, the cooling of a flare loop is initially dominated by conductive losses of heat flux from the corona, which in turn drive an enthalpy flow from the chromosphere to fill the coronal loop \citep{Antiochos_1978}. As the loop temperature decreases and density increases, radiation finally takes over, with the evaporated materials falling back to compensate for the pressure imbalance caused by the inhomogeneous radiative losses \citep{Antiochos_1980}. For a loop with its cross section expanding with height, the reverse cross-sectional contraction toward the loop base acts like a bottle neck to block the downward transportation of both energy and mass, {which would significantly alter the thermodynamic evolution of the loop.}
 
During the conductive-dominated cooling stage, the {upward cross-sectional expansion (or downward cross-sectional contraction)} suppresses the conductive losses from the corona, consequently leading to an elongation of the conductive cooling timescale {compared with the case of a uniform cross section (see Appendix \ref{appB1}). The suppression of the downward thermal conduction meanwhile results in a reduction of total materials evaporated into the corona, even though the conductive-dominated cooling now lasts for a longer time. Therefore, the loop density (temperature) at the start of the radiative-dominated cooling becomes lower (higher) than it would be for a uniform cross section, which further elongates the radiative cooling timescale (see Equation (\ref{taur})). Combining both factors together, the loop will experience a longer-than-expected overall cooling. Adopting the areal function presented in Appendix \ref{appB1} (Equation (\ref{arealfun})) to characterize the loop cross-sectional expansion, and taking advantage of our generalized loop cooling time formula given in Appendix \ref{appB3} (Equation (\ref{tcoold})), an expansion factor of 3 as revealed for the representative late-phase loop will lead to an increase of $\sim$20\% in the overall cooling time, which, with the uncertainty in loop length estimation taken into account, can largely mitigate the discrepancy between the ``theoretical" (by assuming a uniform cross section) and observed cooling times. It is worth pointing out that our analytical consideration of the loop cooling time relies on a parameterized areal function explicitly dependent on temperature, which is somewhat incompatible with those more physically reasonable functions as inferred from observations.}
 
{In spite of a relatively lower loop density at the transition point, such a non-uniform cross section effectively slows down the draining of loop materials backward from the corona during the most time of the radiative-dominated cooling stage until a catastrophic drop occurs (see Appendix \ref{appB2}). At an intermediate temperature (e.g., $\sim$3 MK, the formation temperature of warm coronal emissions) through which the radiative-dominated cooling has been well developed, the loop density could still maintain a high enough level, whereas it would be significantly depleted in case of a uniform cross section. Quantifying the influence of cross-sectional expansion on the mass circulation in a flare loop is beyond the scope of this work. We believe that a numerical survey will help clarify this issue.}

Owing to their longer lengths, the late-phase loops should bear a more notable cross-sectional expansion than the main-phase loops. Reflected to the observations, therefore, we observe a later-than-expected occurrence of the warm coronal late phase (with respect to the case of a uniform cross section), together with an enduring plateau level of the loop EM ($\propto n^2$, see Figure \ref{fig:em}(c)), which may make an additional contribution to the extremely large late-phase peak besides the intense early heating. In passing, we note that such observed patterns have recently been numerically verified using sophisticated radiative hydrodynamic simulations \citep{Reep_2023,Reep_2024}.

In summary, we study the heating and cooling history of a non-eruptive solar flare. The extremely large EUV late phase seen in this flare is mainly due to an intense early heating driven by an erupted-but-failed MFR, and for the first time, we propose that the {extended} cooling of the late-phase loops might arise from an expansion of their cross sections rather than an additional heating.\\

%\begin{acknowledgments}
{We are grateful to the anonymous referee whose insightful comments and suggestions led to a significant improvement of the manuscript.} This work was supported by National Natural Science Foundation of China under grant 12127901. Y.D. is also sponsored by National Key R\&D Program of China under grants 2019YFA0706601 and 2020YFC2201201, as well as Frontier Scientific Research Program of Deep Space Exploration Laboratory under grant 2022--QYKYJH--HXYF--015. \emph{SDO} is a mission of NASA's Living With a Star (LWS) program. J.H.G. is supported by the fellowship of China National Postdoctoral Program for Innovative Talents under Grant Number BX20240159.\\
%\end{acknowledgments}

\appendix

{\section{Validation of the DEM Inversion Results with EVE Data}\label{appA}}
{To verify the validity of the DEM inversion results based on AIA observations (in Section \ref{sec:cool}), we use the inverted DEM distributions to synthesize light curves of several EVE lines, and compare them with the real EVE observations. At each moment, we construct the volume DEM ($DEM_V(T)$) of the flaring region by combining the DEM distributions of all pixels inside the main-phase and late-phase regions. Then we forward synthesize the irradiance of a specific EVE line by
\begin{equation}
I_{\mathrm{line}}=\frac{1}{R^2}\int\!\!\!\int\epsilon(\lambda,T)DEM_V(T)\,dTd\lambda,
\end{equation}
where $R$ is the Sun-Earth distance, and $\epsilon(\lambda,T)$ is the emissivity response matrix as a function of both wavelength $\lambda$ and temperature $T$, which is calculated with the CHIANTI atomic database \citep{Dere_97,delZanna_21}. To account for the instrumental broadening of EVE, the response matrix is further convolved with a Gaussian smoothing function of a 0.7 {\AA} FWHM \citep[cf.][]{Warren_13}.}

\begin{figure}
\epsscale{0.9}
\plotone{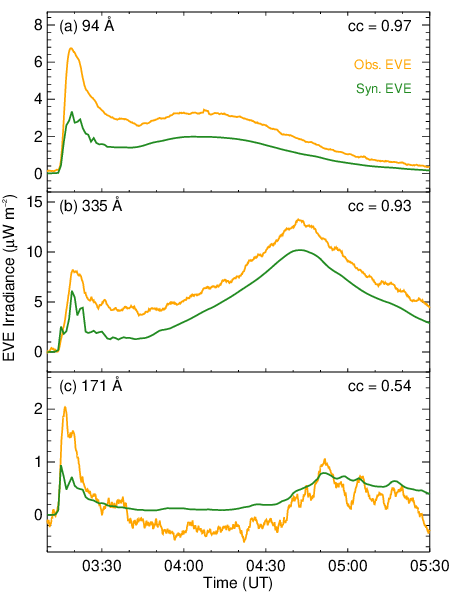}
\caption{Synthetic and observed EVE light curves in several lines (discriminated by the colors). The synthetic curves are derived from the AIA-based DEM inversion results for the flaring region, and the observed curves are displayed the same as in Figure \ref{fig:eve}. In each panel, the cross-correlation coefficient between the two curves is calculated and labeled in the legend. All the time profiles are plotted with the corresponding background levels subtracted. 
 }\label{figappend1}
\end{figure}

{Figure \ref{figappend1} illustrates the comparison between the synthetic and observed EVE light curves in the lines of 94, 335, and 171 {\AA}, respectively. In EVE 94 and 335 {\AA}, the synthetic light curves well characterize the temporal evolution of the observed ones, with the cross-correlation coefficients between them reaching values of over 0.93. In EVE 171~{\AA}, the correlation is not so high, implying an evolving background originating from the cool bulk corona outside the flaring region.}

{For EVE 94 and 335 {\AA}, it is also noted that the synthetic line irradiance is systematically lower than the observed value, although the discrepancy between them is limited within a factor of 1.5 in most of the time.  We attribute this systematic discrepancy to two main factors. The first one comes from our algorithm of region identification and extraction. In identifying the flaring region, we filter out a lot of patchy flaring pixels, which could make a non-negligible contribution to the Sun-as-a-star irradiance as observed by EVE. Second, there could be an inter-instrument calibration coefficient between AIA and EVE, which is not considered when we solely use the AIA data to conduct the DEM inversion. In addition, the uncertainties in the CHIANTI atomic database can also affect the line irradiance synthesis. Nevertheless, considering the narrow discrepancy much less than an order of magnitude, the match between the synthetic and observed EVE light curves is satisfactory, further consolidating the validity of the DEM inversion results in this study.}

\section{Analytical Consideration of the Flare Loop Cooling}\label{appB}
\subsection{Conductive Cooling}\label{appB1}
\citet{Antiochos_1978} studied the conductive cooling accompanied with flow evaporation in a flare loop. Here, we deduce semi-analytical solutions of the model by adopting an explicitly temperature-dependent areal function for the loop cross section. 

Assuming that radiation is negligible and any flows are subsonic, the one-dimensional loop-aligned hydrodynamic equations for the conductive cooling stage are simplified as 
\begin{eqnarray}
&\displaystyle \frac{\partial n}{\partial t}=-\frac{1}{A}\frac{\partial}{\partial s}(Anv), \\
&\displaystyle  p=p_0=\mathrm{const.},  \\
&\displaystyle \frac{1}{A}\frac{\partial}{\partial s}\left[A\left(\frac{\gamma}{\gamma-1}pv-\kappa_0T^{5/2}\frac{\partial T}{\partial s}\right)\right]=0,
\end{eqnarray}
where $t$ and $s$ are the time and loop-aligned coordinate (measured upward from the loop base), $n$, $p$, $T$, and $v$ are the loop density, pressure, temperature, and velocity of the bulk flow, $A$ is the loop cross-sectional area, $\gamma$ is the ratio of specific heat, and $\kappa_0$ is the classical Spitzer conductivity for thermal conduction. Since no energy is drained away (by radiation) from the loop, the pressure keeps a constant level of $p_0$ throughout this stage.

Using the equation of state for fully ionized plasma $p=2nk_BT$ (with $k_B$ being the Boltzmann constant) to relate the loop properties, the combination of the above equations yields
\begin{equation}
\frac{\gamma p_0}{(\gamma-1)y_c}\frac{\partial y_c}{\partial t}=\frac{\kappa_0}{A}\frac{\partial}{\partial s}\left(A\frac{\partial y_c}{\partial s}\right)-\frac{2\kappa_0}{7y_c}
\left(\frac{\partial y_c}{\partial s}\right)^2,\label{singleqnc}
\end{equation}
where we define a temperature-related auxiliary variable $y_c$  as
\begin{equation}
y_c=T^{7/2}.
\end{equation}

Equation (\ref{singleqnc}) is amenable to analytical solutions by separation of variables, i.e.,  
\begin{equation}
y_c(s,t)=T_0^{7/2}\psi(s)\theta_c(t),
\end{equation}
where $T_0$ is the loop-top temperature at the start of the conductive cooling. Therefore, the equation reduces to
\begin{equation}
\frac{\gamma p_0}{(\gamma-1)\kappa_0T_0^{7/2}\theta_c^2}\frac{d\theta_c}{dt}=\frac{1}{A}\frac{d}{ds}\left(A\frac{d\psi}{ds}\right)-\frac{2}{7\psi}\left(\frac{d\psi}{ds}\right)^2,\label{sepeqnc}
\end{equation}
where we can set each side equal to a constant $-(k/L)^2$ that is independent of both $t$ and $s$. In this way, Equation (\ref{sepeqnc}) is separated into two equations
\begin{equation}
\frac{d}{dt}\left(\theta_c^{-1}\right)=\frac{(\gamma-1)k^2\kappa_0T_0^{7/2}}{\gamma p_0L^2}=\frac{1}{\tau_{c0}}\label{timeeqn}
\end{equation}
and
\begin{equation}
\frac{d^2\psi}{ds^2}+\frac{1}{A}\frac{dA}{ds}\frac{d\psi}{ds}-\frac{2}{7\psi}\left(\frac{d\psi}{ds}\right)^2=-\left(\frac{k}{L}\right)^2,\label{spaceeqn}
\end{equation}
where
\begin{equation}
\tau_{c0}=\frac{\gamma p_0L^2}{(\gamma-1)k^2\kappa_0T_0^{7/2}}\label{tauck}
\end{equation}
is the cooling timescale at the start of the conductive cooling.

Equation (\ref{timeeqn}) can be directly integrated to give
\begin{equation}
\theta_c(t)=\left(1+\frac{t}{\tau_{c0}}\right)^{-1},
\end{equation}
and Equation (\ref{spaceeqn}) can be converted into a first-order ordinary differential equation of $(d\psi/ds)^2$ with respect to $\psi$ by assuming an explicitly temperature-dependent form of $A$. Applying the loop-top boundary condition of $d\psi/ds=0$ at $\psi=1$, the integration of Equation (\ref{spaceeqn}) yields
\begin{equation}
\frac{d\psi}{ds}=\frac{k\psi^{2/7}}{AL}\left(2\int_{\psi}^1\psi^{-4/7}A^2\,d\psi\right)^{1/2}.\label{dpsids0}
\end{equation}

To incorporate a loop cross-sectional expansion with height (temperature),  we specify an areal function of 
\begin{equation}
A=A_b\left[1+(\Gamma^2-1)\psi^{12/7}\right]^{1/2},\label{arealfun}
\end{equation}
where $A_b$ is the cross-sectional area at the loop base, and $\Gamma$ is the ratio of cross-sectional areas between the loop apex and base. Inserting it to Equation (\ref{dpsids0}) gives
\begin{equation}
\begin{aligned}
\frac{d\psi}{ds}&=\left(\frac{14}{15}\right)^{1/2}\frac{k}{L}\psi^{2/7}\left\{\frac{(\Gamma^2+4)-\psi^{3/7}\left[5+(\Gamma^2-1)\psi^{12/7}\right]}{1+(\Gamma^2-1)\psi^{12/7}}\right\}^{1/2}\\
&=\frac{k}{Lf(\psi)},\label{dpsids}
\end{aligned}
\end{equation}
where the integrand function $f(\psi)$ is defined as
\begin{equation}
f(\psi)=\left(\frac{15}{14}\right)^{1/2}\psi^{-2/7}\left\{\frac{1+(\Gamma^2-1)\psi^{12/7}}{(\Gamma^2+4)-\psi^{3/7}\left[5+(\Gamma^2-1)\psi^{12/7}\right]}\right\}^{1/2}.
\end{equation}

Equation (\ref{dpsids}) should nevertheless be integrated numerically (except for $\Gamma=1$), which gives
\begin{equation}
\frac{ks}{L}=\int_0^{\psi}f(\psi)\,d\psi=I_{f}(\psi),\label{tprofk}
\end{equation}
where we make the identification $I_{f}(\psi)=\int_0^{\psi}f(\psi)d\psi$ for simplicity. Applying the boundary condition of $\psi=1$ at $s=L$, the eigenvalue of $k$ is then determined by the  definite integral of
\begin{equation}
k=\int_0^1f(\psi)\,d\psi=I_{f}(1).\label{eigenvaluek}
\end{equation}

With the determined $k$, the conductive cooling timescale of the loop (from Equation (\ref{tauck})) is now expressed as
\begin{equation}
\tau_{c0}=\frac{\gamma p_0L^2}{(\gamma-1)I_{f}^2(1)\kappa_0T_0^{7/2}},\label{taucdelt}
\end{equation}
and the temperature profile of the loop (from Equation (\ref{tprofk})) is implicitly formulated as
\begin{equation}
\frac{s}{L}=\frac{I_{f}(\psi)}{I_{f}(1)}.
\end{equation}
Combining them together, the loop solution is finally given as
\begin{equation}
T(s,t)=T_0\psi^{2/7}\left(1+\frac{t}{\tau_{c0}}\right)^{-2/7}.\label{soltc}
\end{equation}

\begin{figure}
\epsscale{0.9}
\plotone{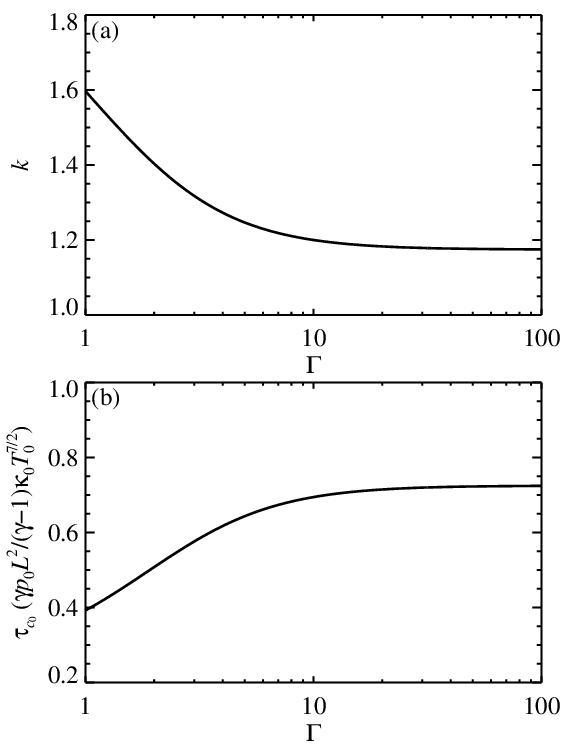}
\caption{Variation of the eigenvalue $k$ (a) and conductive cooling time $\tau_c$ (b) vs.\ the expansion factor $\Gamma$ according to the semi-analytical solution for conductive cooling (Equations (\ref{eigenvaluek}) and (\ref{taucdelt})). }\label{figappend2}
\end{figure}

\begin{figure*}
\epsscale{0.7}
\plotone{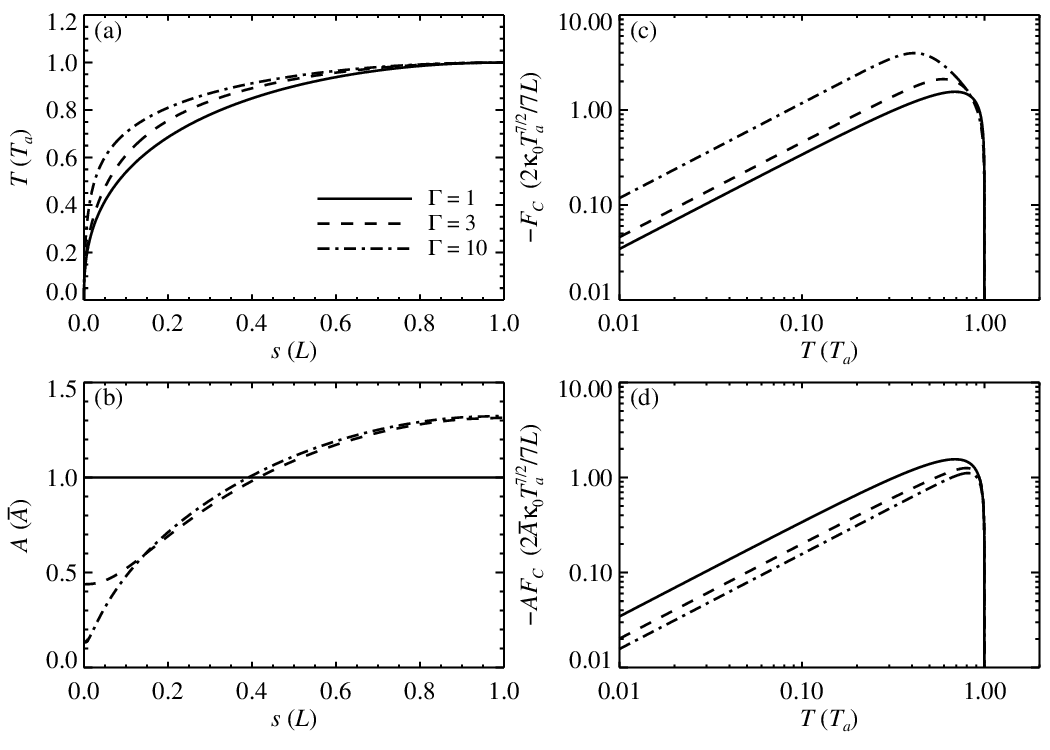}
\caption{Profiles of the loop temperature (a), cross-sectional area (b), heat flux (c), and total conductive losses (d) based on the semi-analytical solutions for conductive cooling. The curves are calculated with different values of the expansion factor (discriminated by different line styles).}\label{figappend3}
\end{figure*}

Figure \ref{figappend2} displays the variation of $k$ and $\tau_{c0}$ versus $\Gamma$.  It is seen that with the increase of $\Gamma$, $k$ decreases and then $\tau_{c0}$ increases. Therefore, a prolonging of the conductive cooling by loop cross-sectional expansion is illustratively corroborated. To further explore the underlying physics, in Figure \ref{figappend3} we plot the profiles of some loop properties for different values of $\Gamma$ according to our semi-analytical solutions. As the expansion factor increases, the temperature profile becomes less and less rounded (Figure \ref{figappend3}(a)), which leads to an elevation of the heat flux ($F_C=-\kappa_0T^{5/2}dT/ds$) transported downward to the TR (Figure \ref{figappend3}(c)). However, the enhancement of the heat flux is not strong enough to compensate for the decrease of the cross-sectional area toward the loop base (Figure \ref{figappend3}(b)), so that the total conductive losses ($AF_C$) from the corona instead decrease with the increase of $\Gamma$ (Figure \ref{figappend3}(d), where $A$ is normalized by the average cross-sectional area of the loop $\bar{A}=\int_0^LA\,ds/L$). Since thermal conduction is the only way to drain energy from the corona during this stage, the decrease of the conductive losses by loop cross-sectional expansion consequently results in a longer conductive cooling time.

\subsection{Radiative Cooling} \label{appB2}
\citet{Antiochos_1980} and \citet{Cargill_1995} theoretically investigated the radiative cooling accompanied with flow drainage in a flare loop. Here, we construct analytical solutions of the cooling loop, which were not fully addressed in their original papers. 

Assuming that the contribution of thermal conduction is marginal and any flows along the loop are subsonic, the governing equations for the radiative cooling stage are simplified as
\begin{eqnarray}
&\displaystyle \frac{\partial n}{\partial t}=-\frac{1}{A}\frac{\partial}{\partial s}(Anv),  \\
&\displaystyle  \frac{\partial p}{\partial s} = 0,  \\
&\displaystyle \frac{\partial}{\partial t}\left(\frac{p}{\gamma-1}\right)=-\frac{1}{A}\frac{\partial}{\partial s}\left(\frac{\gamma}{\gamma-1}Apv\right)-n^2\Lambda(T),
\end{eqnarray}
where  $\Lambda(T)$ is the optically thin radiative loss function.  For analytical convenience, $\Lambda(T)$ is simplified as a single power-law form of $\Lambda(T)=\chi T^{-l}$, with $\chi$ and $l\ (l>0)$ being two constants.

Combining the above equations yields a single equation
\begin{widetext}
\begin{equation}
\begin{aligned}
&\frac{(\gamma-1)\chi p}{4\gamma k_B^2}\left[\frac{1}{y_r}\left(\frac{\partial y_r}{\partial s}\right)^2+(l+2)A\frac{\partial}{\partial s}
\left(\frac{1}{A}\frac{\partial y_r}{\partial s}\right)\right]+A\frac{\partial y_r}{\partial t}\frac{\partial }{\partial s}\left(\frac{1}{A}\frac{\partial y_r}{\partial s}\right)
-\frac{\partial^2y_r}{\partial t\partial s}\frac{\partial y_r}{\partial s}\\
&\ \ \ +\frac{1}{\gamma p}\frac{dp}{dt}\left\{\big[(\gamma-1)(l+2)+1\big]\left(\frac{\partial y_r}{\partial s}\right)^2-(\gamma-1)(l+2)
Ay_r\frac{\partial}{\partial s}\left(\frac{1}{A}\frac{\partial y_r}{\partial s}\right)\right\}=0,\label{singleqn}
\end{aligned}
\end{equation}
\end{widetext}
where we introduce another temperature-related auxiliary variable $y_r$ defined as
\begin{equation}
y_r=T^{l+2}.
\end{equation}

Equation (\ref{singleqn}) is also amenable to analytical solutions by separation of variables, i.e.,  
\begin{equation}
y_r(s,t)=T_{*}^{l+2}\xi(s)\theta_r(t),\label{yst}
\end{equation}
and
\begin{equation}
p(t)=p_{*}\phi_r(t),\label{pt}
\end{equation}
where $T_{*}$ and $p_{*}$ are the loop-top temperature and pressure at the start of the radiative cooling. Putting Equations (\ref{yst}) and (\ref{pt}) into Equation (\ref{singleqn}) gives
\begin{widetext}
\begin{equation}
\begin{aligned}
& \frac{1}{\xi}\left(\frac{d\xi}{ds}\right)^2+(l+2)A\frac{d}{ds}\left(\frac{1}{A}\frac{d\xi}{ds}\right)
+\frac{\gamma\tau_{r*}}{\phi_r}\frac{d\theta_r}{dt}\left[A\xi\frac{d }{ds}\left(\frac{1}{A}\frac{d\xi}{ds}\right)-\left(\frac{d\xi}{ds}\right)^2\right]\\
&\ \ \ +\frac{\tau_{r*}\theta_r}{\phi_r^2}\frac{d\phi_r}{dt}\left\{\big[(\gamma-1)(l+2)+1\big]\left(\frac{d\xi}{ds}\right)^2-(\gamma-1)(l+2)
A\xi\frac{d}{ds}\left(\frac{1}{A}\frac{d\xi}{ds}\right)\right\}=0,\label{sepeqn}
\end{aligned}
\end{equation}
\end{widetext}
where
\begin{equation}
\tau_{r*}=\frac{4k_B^2T_{*}^{l+2}}{(\gamma-1)\chi p_{*}}\label{taur}
\end{equation}
is the cooling timescale at the start of the radiative cooling.

Further specifying time-dependent forms of $\theta_r(t)$ and $\phi_r(t)$ as
\begin{equation}
\theta_r(t)=(1+\eta t)^{-\nu}\label{thetat}
\end{equation}
and
\begin{equation}
\phi_r(t)=(1+\eta t)^{-(\nu+1)},\label{phit}
\end{equation}
where $\eta$ and $\nu$ are two constants, the integration of Equation (\ref{sepeqn}) with respect to $s$ yields
\begin{equation}
\frac{d\xi}{ds}=\frac{A(s)}{A_b}\left(\frac{d\xi}{ds}\right)_b\left(\frac{\xi}{\xi_b}\right)^{-1/(l+2)}\left(\frac{\xi+\mu}{\xi_b+\mu}\right)^g,\label{eqnxi1}
\end{equation}
where we make the identifications
\begin{equation}
\mu=\frac{l+2}{\Big\{\big[(\gamma-1)(l+1)-1)\big]\nu+(\gamma-1)(l+2)\Big\}\eta\tau_{r*}}
\end{equation}
and
\begin{equation}
g(\nu)=\frac{1}{l+2}+\frac{(\gamma-1)(l+1)\nu+(\gamma-1)(l+2)+1}{\big[(\gamma-1)(l+1)-1)\big]\nu+(\gamma-1)(l+2)},\label{gnu}
\end{equation}
for simplicity, and the subscript $b$ denotes quantities evaluated at the loop base.

Applying the boundary condition of  $d\xi/ds=0$ at $\xi=1$, it turns out that $\mu=-1$. Therefore,
\begin{equation}
\eta(\nu)=-\frac{l+2}{\Big\{\big[(\gamma-1)(l+1)-1)\big]\nu+(\gamma-1)(l+2)\Big\}\tau_{r*}},\label{etanu}
\end{equation}
and Equation (\ref{eqnxi1}) can be rewritten into
\begin{equation}
\frac{d\xi}{ds}=CA(s)\xi^{-1/(l+2)}(1-\xi)^{g},\label{eqnxi2}
\end{equation}
where we define an integral constant
\begin{equation}
C=\frac{1}{A_b}\left(\frac{d\xi}{ds}\right)_b\frac{\xi_b^{1/(l+2)}}{(1-\xi_b)^{g}}
\end{equation}
for simplicity.

The strong dependence of $\xi$ on $T$ ($\xi\sim T^{l+2}$) guarantees that $\xi_b\ll0$. Hence, it is reasonable to set $\xi_b=0$ for mathematical convenience. In this way, the integration of Equation (\ref{eqnxi2}) yields
\begin{equation}
\beta\left(\xi;\frac{l+3}{l+2},1-g\right)=CI_A(s),
\end{equation}
where $\beta$ is the incomplete Beta function, and  the integral $I_A(s)$ is defined as
\begin{equation}
I_A(s)=\int_0^sA(s)\,ds.
\end{equation}

Applying the boundary condition of $\xi=1$ at $s=L$ to determine the integral constant, it is found that
\begin{equation}
C=\frac{1}{\bar{A}L}B\left(\frac{l+3}{l+2},1-g\right),\label{intconst}
\end{equation}
where $B$ is the complete Beta function. With the determined $C$, the temperature profile of the loop is implicitly given by
\begin{equation}
\beta_r\left(\xi;\frac{l+3}{l+2},1-g\right)=\frac{I_A(s)}{\bar{A}L},
\end{equation}
where $\beta_r$ is the regularized incomplete Beta function (i.e., $\beta_r=\beta/B$).

The functional forms of Equations (\ref{thetat}) and (\ref{phit}) imply a $T$--$n$ scaling relation of $T(t)\sim n(t)^{\alpha}$ during the radiative cooling stage, with the index $\alpha$ satisfying 
\begin{equation}
-\frac{\nu}{l+2}=\alpha\left[-(\nu+1)+\frac{\nu}{l+2}\right],
\end{equation}
from which $\nu$ can be expressed as a function of $\alpha$ as
\begin{equation}
\nu=-\frac{(l+2)\alpha}{(l+1)\alpha-1}.
\end{equation}
Using this relation to substitute $\nu$ in Equations (\ref{gnu}) and (\ref{etanu}), it is found that
\begin{equation}
g(\alpha)=\frac{(l+2)(\alpha-\gamma+1)-\gamma}{(l+2)(\alpha-\gamma+1)}
\end{equation}
and
\begin{equation}
\eta(\alpha)=-\frac{(l+1)\alpha-1}{(\alpha-\gamma+1)\tau_{r*}}.\label{etaalpha}
\end{equation}

With a monotonic increase of the temperature from the loop base, a temperature maximum occurring at the loop apex requires that $g(\alpha)>0$ (from Equation (\ref{eqnxi2})). Meanwhile, a valid definition of the Beta function in Equation (\ref{intconst}) requires that $1-g(\alpha)>0$. Combining these two conditions gives $0<1-g(\alpha)<1$, which imposes a constraint on the evaluation of $\alpha$ as
\begin{equation}
\alpha>\gamma-1+\frac{\gamma}{l+2}.
\end{equation}
The lower bound of $\alpha$, $\gamma-1$ (for $l\rightarrow\infty$), corresponds to an adiabatic expansional drainage of loop material from the corona. In passing, it is readily verified that $\eta(\alpha)<0$ for any positive values of $l$ as long as $\gamma>4/3$ (naturally satisfied for the solar coronal circumstance where $\gamma$ is typically set to be 5/3).

Using $\alpha$ to construct the loop solutions, it is found that
\begin{equation}
T(s,t)=T_{*}\xi^{1/(l+2)}(1+\eta t)^{\alpha/[(l+1)\alpha-1]},\label{soltr}
\end{equation}
\begin{equation}
p(t)=p_{*}(1+\eta t)^{(\alpha+1)/[(l+1)\alpha-1]},\label{solp}
\end{equation}
and
\begin{equation}
n(s,t)=\frac{p_{*}}{2k_BT_{*}}\xi^{-1/(l+2)}(1+\eta t)^{1/[(l+1)\alpha-1]}.\label{soln}
\end{equation}
With the obtained solutions, the flow velocity can be calculated from the equation of continuity and formulated as 
\begin{equation}
\begin{aligned}
v(s,t)&=\left[\frac{(\gamma-1)}{\gamma p}\frac{dp}{dt}-\frac{1}{T}\frac{\partial T}{\partial t}-\frac{(\gamma-1)\chi p}{4\gamma k_B^2}T^{-(l+2)}\right]\left(\frac{1}{T}\frac{\partial T}{\partial s}\right)^{-1}\\
&=-\frac{l+2}{\gamma}\left[B\left(\frac{l+3}{l+2},1-g\right)\right]^{-1}\frac{\bar{A}L}{A(s)\tau_{r*}}\\
&\  \times\xi^{1/(l+2)}\left(1-\xi\right)^{1-g}(1+\eta t)^{-1},
\end{aligned}
\end{equation}
and then the total mass drainage $\mathcal{F}=Anv$ is expressed as
\begin{equation}
\begin{aligned}
\mathcal{F}(s,t)&=-\frac{l+2}{2\gamma }\left[B\left(\frac{l+3}{l+2},1-g\right)\right]^{-1}
\frac{\bar{A}p_{*}L}{k_BT_{*}\tau_{r*}}\\
&\ \ \times\left(1-\xi\right)^{1-g}(1+\eta t)^{-1+1/[(l+1)\alpha-1]}.\label{solf}
\end{aligned}
\end{equation}

\begin{figure*}
\epsscale{0.7}
\plotone{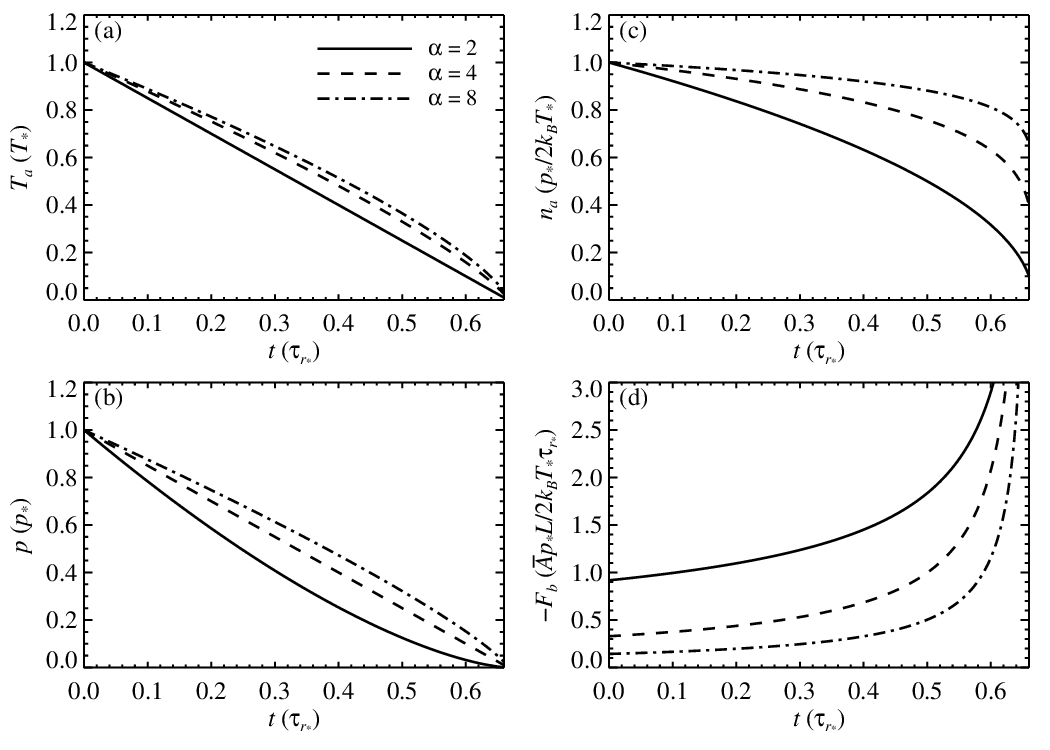}
\caption{Temporal evolutions of the loop-top temperature (a), pressure (b), density (c), and mass flow through the loop base (d) based on the analytical solutions for radiative cooling (Equations (\ref{soltr})--(\ref{soln}) and (\ref{solf})). The curves are calculated with commonly adopted values of $\gamma$ and $l$ at 5/3 and 1/2, respectively. Different line styles denote cases of different values of $\alpha$ in the $T$--$n$ scaling relation (indicated in the legend).}\label{figappend4}
\end{figure*}

Figure \ref{figappend4} shows the temporal evolutions of the loop-top temperature, pressure, density, and total mass drainage through the loop base according to our derived analytical solutions. Here, we set $\gamma$ and $l$ at their commonly adopted values of 5/3 and 1/2, respectively \citep[e.g.,][]{Bradshaw_2010}. With these parameters, the case of $\alpha=2$ refers to a loop of a constant cross section, as validated from previous numerical simulations \citep{Serio_1991,Jakimiec_1992}. For a loop with its cross section expanding with height, nevertheless, the reverse contraction of cross section toward the loop base would effectively block the downward mass drainage from the corona (similar to the effect of the blockage of conductive energy losses during the conductive cooling stage), which leads to an increase of $\alpha$ in the $T$--$n$ scaling relation (the loop cools mainly by radiation rather than enthalpy losses). As shown in Figures \ref{figappend4}(a)--(c), the increase of $\alpha$ barely affects the overall radiative cooling time, since the negative valued cooling rate $\eta$ itself just has a very weak or even zero dependence on $\alpha$ for typical values of $\gamma$ and $l$. According to Equations (\ref{soltr})--(\ref{soln}), nevertheless, it reduces the positive powers on the time-dependent factor $1+\eta t$, hence making the decay curves more convex (or less concave). In this way, the loop properties can maintain a high enough level for a longer time (especially for $n$) until a catastrophic drop occurs.  Meanwhile, the increase of $\alpha$ also elevates the value of the Beta function in Equation (\ref{solf}) by decreasing the argument $1-g$. As an a posteriori verification, the initial level of mass drainage from the corona is indeed found to be suppressed for an upward cross-sectional expansion compared with the case of a constant cross section (see Figure \ref{figappend4}(d)). 

\subsection{Overall Cooling Time} \label{appB3}
 \citet{Cargill_1995} analytically estimated the overall cooling time of a flare loop by assuming that conduction dominates initially and radiation takes over later on. Here, we improve their simple cooling model. At the transition point from conductive cooling to radiative cooling, the two cooling rates should be equal, which, by relating Equations (\ref{soltc}) and (\ref{soltr}), leads to
\begin{equation}
\tau_{c*}=\frac{2(\alpha-\gamma+1)}{7\alpha}\tau_{r*},\label{taucr}
\end{equation}
where $\tau_{c*}$ is the instantaneous conductive cooling timescale evaluated at the transition point. Note that in \citet{Cargill_1995}, it is simply set that $\tau_{c*}=\tau_{r*}$.

The evaluation of $\tau_{c*}$ and $\tau_{r*}$ needs a determination of the transition temperature $T_{*}$, a quantity relating them to the corresponding initial values $\tau_{c0}$ and $\tau_{r0}$ by
\begin{equation}
\tau_{c*}=\left(\frac{T_{*}}{T_0}\right)^{-7/2}\tau_{c0} \label{taucc}
\end{equation}
and
\begin{equation}
\tau_{r*}=\left(\frac{T_{*}}{T_0}\right)^{l+2}\tau_{r0}. \label{taurr}
\end{equation}
Combining Equations (\ref{taucr})--(\ref{taurr}) yields
\begin{equation}
\frac{T_{*}}{T_0}=\left[\frac{7\alpha}{2(\alpha-\gamma+1)}\frac{\tau_{c0}}{\tau_{r0}}\right]^{2/(2l+11)},
\end{equation}
\begin{equation}
\tau_{c*}=\left[\frac{2(\alpha-\gamma+1)}{7\alpha}\right]^{7/(2l+11)}\tau_{c0}^{2(l+2)/(2l+11)}\tau_{r0}^{7/(2l+11)},
\end{equation}
and
\begin{equation}
\tau_{r*}=\left[\frac{7\alpha}{2(\alpha-\gamma+1)}\right]^{2(l+2)/(2l+11)}\tau_{c0}^{2(l+2)/(2l+11)}\tau_{r0}^{7/(2l+11)}.
\end{equation}
Here, we use the condition of $p_{*}=p_0$, since the loop pressure is assumed to keep unchanged during the whole conductive cooling stage.

Using Equation (\ref{soltc}), the conductive cooling time $\tau_{\mathrm{cond}}$ is estimated to be
\begin{equation}
\tau_{\mathrm{cond}}=\left[\left(\frac{T_{*}}{T_0}\right)^{-7/2}-1\right]\tau_{c0}\approx\tau_{c*}.
\end{equation}
Applying Equation (\ref{soltr}), the radiative cooling time $\tau_{\mathrm{rad}}$ is directly written as 
\begin{equation}
\tau_{\mathrm{rad}}=\frac{\alpha-\gamma+1}{(l+1)\alpha-1}\tau_{r*}.
\end{equation}
Summing them together, the overall cooling time $\tau_{\mathrm{cool}}$ is then given by
\begin{widetext}
\begin{equation}
\tau_{\mathrm{cool}}\approx\left[\frac{2(\alpha-\gamma+1)}{7\alpha}\right]^{7/(2l+11)}\left\{1+\frac{7\alpha}{2\big[(l+1)\alpha-1\big]}\right\}\tau_{c0}^{2(l+2)/(2l+11)}\tau_{r0}^{7/(2l+11)}.\label{tcool}
\end{equation}
\end{widetext}

Setting $\gamma=5/3$ and $l=1/2$, Equation (\ref{tcool}) reduces to 
\begin{equation}
\tau_{\mathrm{cool}}\approx\left[\frac{2(3\alpha-2)}{21\alpha}\right]^{7/12}\frac{2(5\alpha-1)}{3\alpha-2}\tau_{c0}^{5/12}\tau_{r0}^{7/12},\label{tcoold}
\end{equation}
whose form is essentially the same as that presented in \citet[][their Equation (13E)]{Cargill_1995} except for a small difference in the proportionality coefficient. (Here the  proportionality coefficient just varies in a narrow range of 1.61--1.71 for $\alpha\ge2$, very close to the value of 5/3 in their original formula.) In spite of the general consistency, our cooling model gives a ratio of $\tau_{\mathrm{rad}}/\tau_{\mathrm{cond}}=7\alpha/(3\alpha-2)>7/3$ rather than the value of $2/3$ in \citet{Cargill_1995}, indicating a dominance of radiative cooling stage in the overall cooling time, and hence being more physically reasonable.

Furthermore, compared with that in \citet{Cargill_1995}, our formalism of the conductive cooling timescale (Equation (\ref{tauck})) is improved by a factor of $\gamma/k^2$.  Inserting the expressions of $\tau_{c0}$ and $\tau_{r0}$ into Equation (\ref{tcoold}), therefore, the complete Cargill's cooling time formula for a uniform cross section (where $\alpha=2$, and $k\approx1.60$ as seen from Figure \ref{figappend2}(a)) is finally modified to
\begin{equation}
\tau_{\mathrm{cool}}\approx2.04\times10^{-2}L^{5/6}T_{0}^{-1/6}n_{0}^{-1/6},
\end{equation}
with the proportionality coefficient reduced by a factor of $\sim15$\% with respect to the original value in \citet[][their Equation (14E)]{Cargill_1995}.

\bibliography{ref.bib}
\bibliographystyle{aasjournal}

\end{document}